\documentclass[%
reprint,
superscriptaddress,
 amsmath,amssymb,
 aps,
 pra,
]{revtex4-1}

\usepackage{graphicx}
\usepackage{dcolumn}
\usepackage{bm}

\usepackage[utf8]{inputenc} 

\usepackage{multirow}
\usepackage[warn]{mathtext}
\usepackage{amssymb}

\usepackage{textcomp} 
\usepackage{indentfirst} 
\usepackage{amsmath} 
\usepackage{graphicx}
\DeclareGraphicsExtensions{.pdf,.png,.jpg}
\usepackage{pgfplots}
\pgfplotsset{compat=1.13}

\usepackage{hyperref}
\hypersetup{
    colorlinks=true,
    linkcolor=blue,
    filecolor=black,      
    urlcolor=black,
    citecolor= violet
}

\usepackage{algpseudocode}

\usepackage{adjustbox}
\usepackage{tabularx}

\renewcommand{\bibname}{References}

\usepackage[nottoc]{tocbibind}

\usepackage{xcolor}
\definecolor{C0}{HTML}{4C72B0}
\definecolor{C1}{HTML}{DD8452}
\definecolor{C2}{HTML}{55A868}
\definecolor{C3}{HTML}{C44E52}
\definecolor{emerald}{HTML}{006400}



\begin{document}

\title{Coupler microwave-activated controlled phase gate on fluxonium qubits}

\author{Ilya~A.~Simakov}
\email{simakov.ia@phystech.edu}
\affiliation{Russian Quantum Center, 143025 Skolkovo, Moscow, Russia}
\affiliation{National University of Science and Technology ``MISIS'', 119049 Moscow, Russia}
\affiliation{Moscow Institute of Physics and Technology, 141701 Dolgoprudny, Russia}

\author{Grigoriy~S.~Mazhorin}
\affiliation{Russian Quantum Center, 143025 Skolkovo, Moscow, Russia}
\affiliation{National University of Science and Technology ``MISIS'', 119049 Moscow, Russia}
\affiliation{Moscow Institute of Physics and Technology, 141701 Dolgoprudny, Russia}

\author{Ilya~N.~Moskalenko}
\altaffiliation[Present address: ]{ Department of Applied Physics, Aalto University, Espoo, Finland}
\affiliation{Russian Quantum Center, 143025 Skolkovo, Moscow, Russia}
\affiliation{National University of Science and Technology ``MISIS'', 119049 Moscow, Russia}

\author{Nikolay~N.~Abramov}
\affiliation{Russian Quantum Center, 143025 Skolkovo, Moscow, Russia}
\affiliation{National University of Science and Technology ``MISIS'', 119049 Moscow, Russia}

\author{Alexander~A.~Grigorev}
\affiliation{National University of Science and Technology ``MISIS'', 119049 Moscow, Russia}

\author{Dmitry~O.~Moskalev}
\affiliation{Dukhov Research Institute of Automatics (VNIIA),  Moscow 127055, Russia}
\affiliation{FMN Laboratory, Bauman Moscow State Technical University, Moscow 105005, Russia}
\author{Anastasiya~A.~Pishchimova}
\affiliation{Dukhov Research Institute of Automatics (VNIIA),  Moscow 127055, Russia}
\affiliation{FMN Laboratory, Bauman Moscow State Technical University, Moscow 105005, Russia}
\author{Nikita~S.~Smirnov}
\affiliation{Dukhov Research Institute of Automatics (VNIIA),  Moscow 127055, Russia}
\affiliation{FMN Laboratory, Bauman Moscow State Technical University, Moscow 105005, Russia}
\author{Evgeniy~V.~Zikiy}
\affiliation{Dukhov Research Institute of Automatics (VNIIA),  Moscow 127055, Russia}
\affiliation{FMN Laboratory, Bauman Moscow State Technical University, Moscow 105005, Russia}
\author{Ilya~A.~Rodionov}
\affiliation{Dukhov Research Institute of Automatics (VNIIA),  Moscow 127055, Russia}
\affiliation{FMN Laboratory, Bauman Moscow State Technical University, Moscow 105005, Russia}
\author{Ilya S. Besedin}
\altaffiliation[Present address: ]{Department of Physics, ETH Zurich, Zurich, Switzerland}
\affiliation{Russian Quantum Center, 143025 Skolkovo, Moscow, Russia}
\affiliation{National University of Science and Technology ``MISIS'', 119049 Moscow, Russia}

\date{\today}

\begin{abstract}

Tunable couplers have recently become one of the most powerful tools for implementing two-qubit gates between superconducting qubits. A tunable coupler typically includes a nonlinear element, such as a SQUID, which is used to tune the resonance frequency of an LC circuit connecting two qubits. 
Here we propose a complimentary approach where instead of tuning the resonance frequency of the tunable coupler by applying a quasistatic control signal, we excite by microwave the degree of freedom associated with the coupler itself. Due to strong effective longitudinal coupling between the coupler and the qubits, the frequency of this transition strongly depends on the computational state, leading to different phase accumulations in different states.
Using this method, we experimentally demonstrate a CZ gate of 44~ns duration on a fluxonium-based quantum processor, obtaining a fidelity of $97.6\pm 0.4 \%$ characterized by cross-entropy benchmarking.

\end{abstract}

\maketitle


\section{Introduction}


During the last decade, there has been a tremendous progress in implementing fluxonium qubits \cite{Manucharyan_2009, PhysRevX.9.041041, PhysRevX.11.011010, moskalenko2019planar} as the building blocks of a superconducting quantum processor. Since their proposal \cite{Manucharyan_2009} fluxoniums have become a perspective experimental platform for creating superconducting quantum circuits with a wide range of control parameters due to their richer energy level structure compared to widely used transmons \cite{Koch_2007, Google_2Q}. Due to its large anharmonicity, the fluxonium qubit has significant advantages over the transmon qubit in terms of coherence times, single-qubit gate fidelities, and leakage rates. Fluxonium circuits implemented in a 2D architecture \cite{PhysRevX.11.011010} showed qubit coherence of times $T_1, T_{2e} \approx 300~\mu s$, while fluxonium qubits installed in 3D cavities demonstrated coherence times higher than 1 ms and average single-qubit gate fidelities above 0.9999 \cite{Somoroff2021MillisecondCI}.


The implementation of two-qubit gates on the fluxonium platform has moved from directly coupled qubits \cite{ficheux2021fast, Bao2022} to a more scalable approach with tunable interaction via a coupler element \cite{Moskalenko_2022}. 
Similar to transmons, two-qubit gates on fluxoniums can be implemented using the flux tunability of qubits.
The obvious disadvantage of tuning the flux is that the coherence time of the qubit is only large when operated at the flux sweet spot. Another disadvantage of flux-activated gates is that sweeping the frequency of the qubits leads to a Landau-Zener-type interaction with strongly coupled two-level defects that would normally be off-resonant with the qubit \cite{Ivakhnenko2023}. This has been identified as a major problem when many two-qubit gates need to be performed \cite{arute2019quantum, Krinner2022}. 

An alternative way of implementing two-qubit gates is by inducing transition interactions between the qubits with microwaves. Microwave-activated gates include transitions via higher-excited energy levels \cite{ficheux2021fast, Xiong2022ARBCZ} or cross-resonance gates \cite{Nesterov2022, Dogan2022}. These gates rely on direct capacitive coupling between the qubits, which has a drawback of large residual ZZ interactions between the qubits when no gate is applied.

\begin{figure*}[t]
    \center{\includegraphics[width=\linewidth]{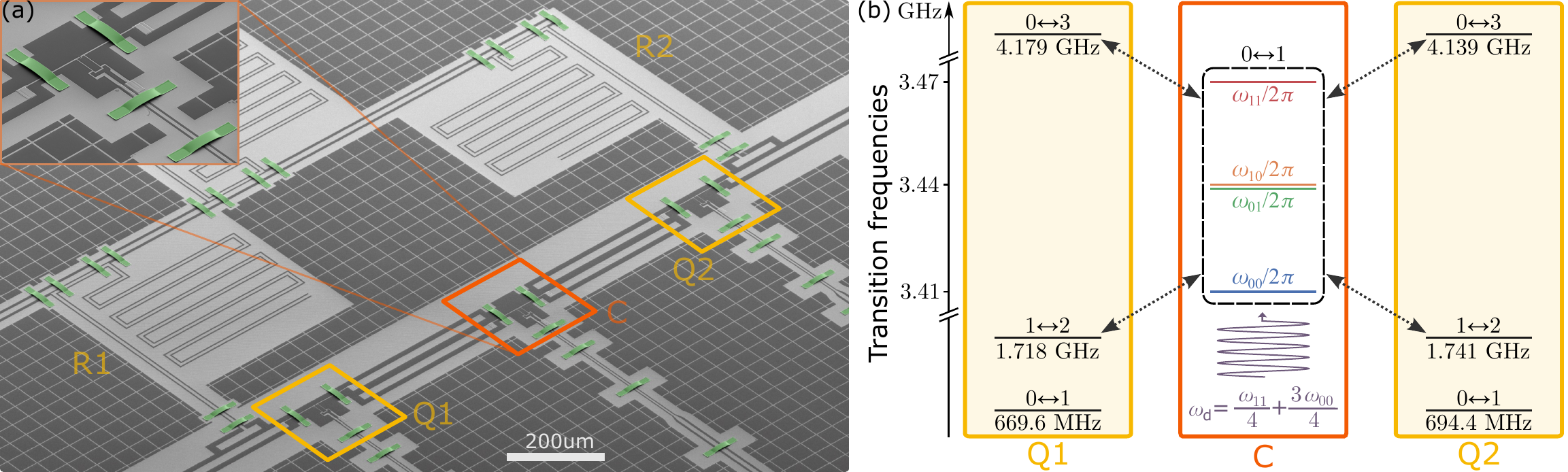}}
    \caption{The device and gate concept. (a) The SEM image of the fluxonuim quantum processor. The computational qubits Q1 and Q2 are capacitively coupled to the individual readout resonators R1 and R2, those in turn are connected to the readout transmission line. 
    Both qubits are also capacitively coupled to the middle fluxonium  C, which enlarged image is shown in the upper left corner. Each fluxonium has its own excitation line. 
    (b) Schematic diagram of the relevant fluxonium state transitions. The arrows show the interaction between the higher fluxoniums energy transitions and the coupler $0-1$ transition, that results in the dispersive shift of the coupler. The frequencies $\omega_{ij}$ correspond to the coupler $0-1$ transitions, associated with the $|ij\rangle$ state of the computational qubits. With the violet color we depicted the external microwave drive on the coupler required for the proposed gate implementation.
    }
    \label{fig:schemelevels}
\end{figure*}

In this paper, we propose a CZ gate scheme between two low-frequency fluxoniums coupled via an extra coupler fluxonium. Instead of applying a magnetic flux to the coupler element to change the coupling strength between the qubits, we activate microwave transitions in the coupler element. Due to strong interaction between this element and the computational qubits, the spectrum of these transitions is state-dependent. As a result, the detuning of the microwave drive depends on computational states, which results in different acquired phases in these states. This gate principle is closely related to the microwave-activated CZ gate~\cite{ficheux2021fast}, as well as the iToffoli gate proposal \cite{Baker2022}. Compared to the former, we obtain the geometric phase by exciting a tunable coupler element degree of freedom, which allows to lower the residual coupling to below~10~kHz. Compared to the latter, we consider not a three-qubit gate, where the qubit directly coupled to two neighbors is driven, but rather re-label the driven qubit as the coupler. This allows to ignore the residual coupling between the driven element and the computational, since the coupler is deexcited when no gate is being performed, significantly simplifying the circuit. We experimentally demonstrate the coupler microwave-activated controlled phase (CMACP) by tuning up a 44~ns-long CZ gate with a cross-entropy benchmarking fidelity of $97.6\pm 0.4 \%$.

\section{Device and theoretical background}

We investigate a quantum device consisting of three capacitively coupled fluxonium qubits, two of which are the computational qubits and the remaining one plays the role of a tunable coupler, see Fig.~\ref{fig:schemelevels}a. Each fluxonium has an additional harmonic mode with a frequency close to 2~GHz. The mutual capacitances between the qubits lead to large transversal couplings between the bare fluxonium and harmonic modes.  The details of the circuit layout and parameters are discussed in detail in \cite{Moskalenko2021, Moskalenko_2022}.

Here, the computational qubits are biased at their flux degeneracy points, and the coupler is at zero flux. At the operating point the frequencies of the computational qubits are $669.6$~MHz and $694.4$~MHz, while the coupler frequency is about $3.4$~GHz. At these flux biases, when the coupler is deexcited, the residual effective ZZ and XX couplings between the computational qubits are weak, $-10$ kHz and 500 kHz, respectively.


We consider the ground and excited states of each fluxonium mode, and a microwave signal applied to the coupler close to its resonance frequency. 
With all fluxoniums being detuned from each other, the transversal couplings result in state dressing and dispersive frequency shifts.
The hybridization of the first coupler state with the higher fluxoniums levels result in the dispersive frequency shifts of the main coupler transition. The shifts for the first and the second qubits are equal respectively to 30 MHz and 26 MHz. 
We schematically show the interaction and the resulted dispersive shift of the coupler $0-1$ transition in Fig.~\ref{fig:schemelevels}b.
Because of the very large frequency difference between the coupler and the qubits we neglect driving of the dressed computational qubit modes. Thus the effective Hamiltonian of the coupled multi-fluxonium system can be expressed as


\begin{figure*}
    \center{\includegraphics[width=\linewidth]{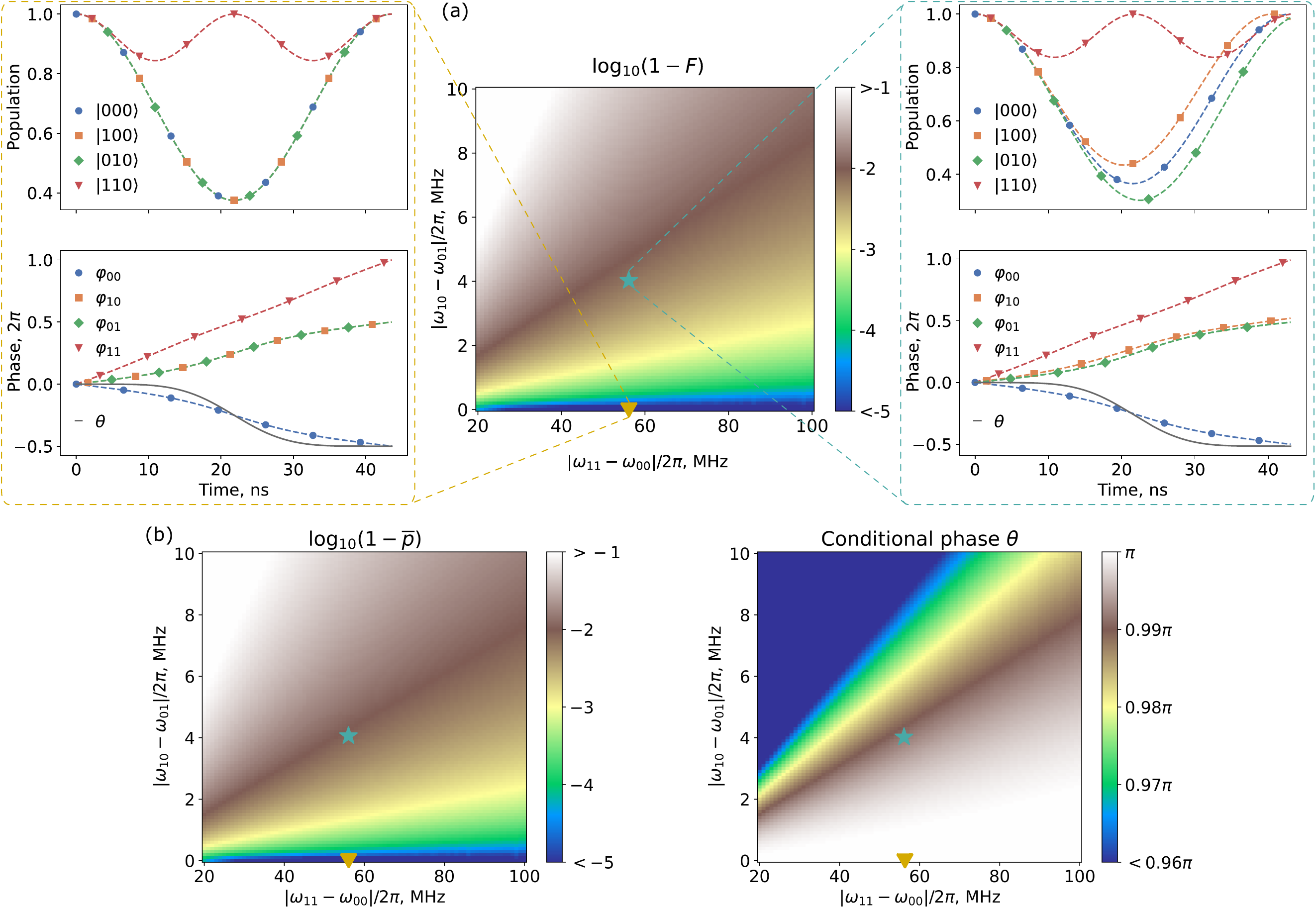}}
    \caption{Gate simulation. (a) The fidelity of a CZ gate $\log_{10}(1-F)$ as a function of the differences between the design-dependent transition frequencies $\omega_{10}$, $\omega_{01}$ and $\omega_{11}$, $\omega_{00}$. The orange triangular correspond to the ideal point when $\omega_{10}=\omega_{01}$ and the left inset plot shows the time evolution of the population and common phase of the four computational states. The dashed gray line shows the acquired phase $\theta = \varphi_{00} - \varphi_{10} - \varphi_{01} + \varphi_{11}$. The fidelity of the such gate is exactly 1. The turquoise star corresponded to the frequencies, observed in the experiment (see Table~\ref{tab:frequencies}). The time dynamics of the computational states associated with the ground state of the coupler and the common phases is shown in the right inset figure. The obtained CZ gate fidelity is 0.992. (b) The average population $\log_{10}(1-\overline{p})$ and the accumulated phase $\theta$ of a CZ gate as a function of the differences between the transition frequencies. We mention that the conditional phase in the blue area reaches the values of $0.5\pi$.}
    \label{fig:simulation}
\end{figure*}

\begin{equation}
   \frac{\mathcal{H}}{\hbar} = -\sum_{i=1,2,c}\frac{\omega_i}{2}\sigma_{zi}
   -\sum_{i=1,2}\frac{\zeta_{ic}}{4}\sigma_{zi}\sigma_{zc} + \Omega \sigma_{xc}\cos{\omega_d t},
   \label{eq:H_dr}
\end{equation}
where $\omega_i/2\pi$ is the renormalized frequency of the $i$-th qubit, $\zeta_{ic}$ is the dispersive coupling strength between the computational qubits and the coupler, and $\sigma_{xi}$ and $\sigma_{zi}$ are Pauli operators acting on the lowest two levels of the $i$-th fluxonium, and $\Omega$ is the drive strength. Here we also neglect the direct interaction between the computational qubits since $\zeta_{12} \ll \zeta_{1c}, \zeta_{2c}$.

The Hilbert space of the system can be separated into four non-interacting subspaces corresponding to different states of the computational qubits. We focus on the transitions between the ground and excited states of the coupler in these subsystems. Within a rotating wave approximation in the frame of the drive signal their Hamiltonians can be written as


\begin{equation}
\frac{\mathcal{H}}{\hbar} = -\frac{\Delta_{mn}}{2}\sigma_{zc} + \frac{\Omega}{2}\sigma_{xc},
\end{equation}
where $\Delta_{mn} = \omega_\mathrm{d} - \omega_{mn}$ are the detunings between the drive tone and the effective coupler frequency corresponding to the computational state $mn~\in~\{|00\rangle, |01\rangle, |10\rangle, |11\rangle \}$, 
\begin{equation}
\omega_{mn} = \omega_c + \frac{(-1)^{m}}{2}\zeta_{1c} + \frac{(-1)^{n}}{2}\zeta_{2c}.
\end{equation}

The energy spectrum of the coupler is schematically shown in Fig.~\ref{fig:schemelevels}b.

The solution of the Schrödinger equation in case of a rectangular microwave pulse with the ground state initial condition on the wave function $\psi(0) = |0\rangle$ is
\begin{equation}
    \psi(t) = 
    \begin{pmatrix}
        \frac{i \Delta_{mn}}{\Omega_R} \sin \left( \frac{\Omega_R t}{2} \right) + \cos \left( \frac{\Omega_R t}{2} \right) \\
        - \frac{i \Omega}{\Omega_R} \sin \left( \frac{\Omega_R t}{2} \right)
    \end{pmatrix},
    \label{eq:rabi_solution}
\end{equation}
where $\Omega_R = \sqrt{\Delta_{mn}^2 + \Omega^2}$ is the generalized Rabi frequency. Apart from the population transfer, these oscillations also result in a phase accumulation of the computational state. 
Note that, for an integer number of Rabi oscillations this phase is either $\pi$ or $2\pi$.

Thereby, each subspace has its own common phase, which depends on the signal detuning from the frequency of the corresponding transition.
The idea is to find such parameters of the drive pulse that the coupler returns to the ground state for all computational states and the common phases $\varphi_{mn}$ acquired by the Rabi oscillations of the corresponding transition satisfy the following condition:
\begin{equation}
    \theta = \varphi_{00} - \varphi_{10} - \varphi_{01} + \varphi_{11} = \pi + 2\pi k, \; k \in \mathbb{Z}.
    \label{eq:phase_conditions}
\end{equation}

Let us make an assumption that the transitional frequencies $\omega_{10}$, $\omega_{01}$  are equal: $\omega_{10}=\omega_{01}=(\omega_{11}+\omega_{00})/2$. 
In this case both conditions mentioned above can be exactly satisfied, and a CZ gate can be implemented if we choose the drive frequency $\omega_\mathrm{d}$ at the half sum of the transition $\omega_{00}$ and $\omega_{10}$: $\omega_\mathrm{d} = (\omega_{10}+\omega_{00})/2$, the amplitude $\Omega = \sqrt{\frac{5}{12}} (\omega_\mathrm{10}-\omega_{00})$, and the duration $\tau = {\frac{\sqrt{6} \pi}{ \omega_{10}-\omega_{00}}}$. Such a signal results in a single Rabi oscillation for the $|00\rangle$, $|01\rangle$ and $|10\rangle$ computational states, and two oscillations for the $|11\rangle$ state. The time dynamics of these four populations and the corresponding common phases under the described external signal is shown in left inset figure in Fig.~\ref{fig:simulation}. We calculate the evolution in the rotating wave approximation assuming that the oscillations are independent and isolated from the rest system.

Though the qubits are designed to be identical, the critical currents of the phase slip junctions turn out slightly different, thus, the transition frequencies $\omega_{10}$ and $\omega_{01}$ are not equal. We simulate the evolution of the system for the different $|\omega_{11}-\omega_{00}|$ and $|\omega_{10}-\omega_{01}|$ under the drive with frequency $\omega_\mathrm{d}=\frac{\omega_{11}+3\omega_{00}}{4}$,  amplitude $\Omega = {\frac{\sqrt{5}}{4\sqrt{3}}} (\omega_\mathrm{11}-\omega_{00})$, and duration $\tau = {\frac{2\sqrt{6} \pi}{ \omega_{11}-\omega_{00}}}$. The average population over the four initial states, accumulated conditional phase $\theta$ and fidelity of the obtained CZ gate fidelity calculated on the computational states $|000\rangle$, $|100\rangle$, $|010\rangle$, $|110\rangle$ are shown in Fig~\ref{fig:simulation}. The turquoise star on the plot indicates the point corresponding to the obtained experimental data (see Table~\ref{tab:frequencies}), the time dynamics of the four populations and the acquired phase for these frequencies is also shown in the right inset plot in Fig~\ref{fig:simulation}. The fidelity of the simulated gate for the experimental frequencies is 0.992 and the duration is 43~ns.

Besides, we mention yet another possible method to realize a CZ gate via the first excited state of the coupler. If one drives the coupler fluxonium directly at the frequency of the $|110\rangle - |111\rangle$ transition for the duration of a single Rabi oscillation, that states acquires a phase of $\pi$. If the amplitude of the drive signal is low compared to the frequency detuning $\omega_{11} - \omega_{10}$, then one can ignore the oscillations between the levels $|000\rangle - |001\rangle$, $|100\rangle - |101\rangle$, and $|010\rangle - |011\rangle$ and get a CZ gate with a given accuracy. Unfortunately, the duration of this drive for high-fidelity gate is several times longer than for the previous method. We provide the numerical estimates in the Appendix~\ref{appendix:110-111}. Below focus on the faster detuned-drive CZ gate implementation.

\section{Experimental results}

The setup used in this experiment is based on an earlier work \cite{Moskalenko_2022}, with a number of modifications, described in Appendix~\ref{appendix:exp_setup}. The device characteristics are provided in Appendix~\ref{appendix:device_parameters}. Here we just briefly summarize the techniques used for single-qubit gates, readout and initialization, and then focus on the coupler microwave-activated CZ gate implementation.

The reset procedure is realized via capacitively coupled microwave antennas as dissipation channels. We tune the qubit to the zero flux point, where the longitudinal relaxation rate is an order of magnitude higher, and then return the qubit to the operating point just before the scheduled pulse sequence. The readout is implemented using adjacent resonators with fidelities 0.67 and 0.62 for the first and second computational qubits for simultaneous qubit measurement. We explain such a poor readout fidelity by weak coupling strength between the qubit and corresponding resonator, which is an innate feature of the chip layout. The single qubit gates are generated from $\pi/2$ rotations around axes in the equatorial plane of the Bloch sphere, realized by Gaussian pulses with 13.3~ns duration. The excitation pulses are applied through to flux line of the qubit. Due to the large amplitude of these pulses we apply a phase error compensation with a virtual rotation about the Z axis after every $\pi/2$ pulse. We construct each of the 24 elements of the single-qubit Clifford group from two $\pi/2$ gates and virtual Z rotations (see Appendix~\ref{appendix:clifford_group}). The average fidelity of single-qubit Clifford group gates $F=0.9928~\pm~0.0003$ has been measured with the cross entropy benchmarking method (Fig.~\ref{fig:xeb}).

\begin{figure}[!t]
    \center{\includegraphics[width=\linewidth]{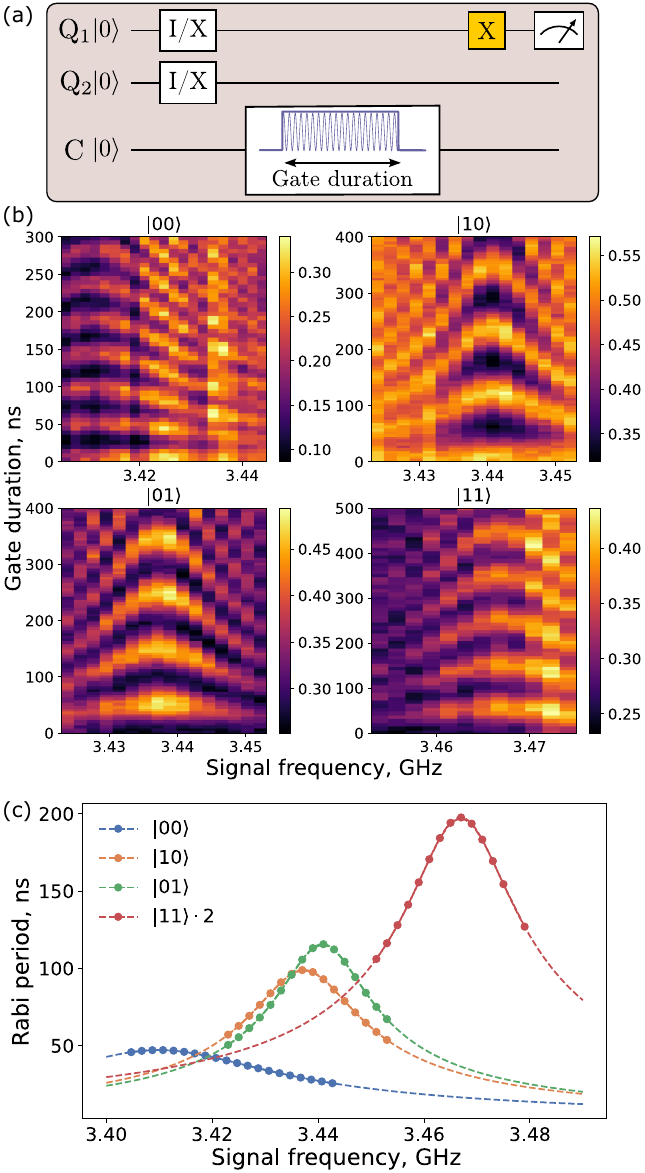}}
    \caption{Rabi oscillations of the coupler for different computational states. (a) The excitation pulse sequence. The first two single qubit gates set the computational qubits into the one of the four computational states. Then we apply the rectangular pulse on the coupling qubit of given duration and finally execute an X gate on the first qubit. We highlight the gate as it is calibrated with the coupler at the ground state. Due to these changes we observe Rabi patterns of the coupler via the computational qubit, shown in fig (b) for the different initial states. (c) The period of coupler Rabi oscillations as a function of the signal frequency obtained from the Rabi patterns. The legend notation $|11\rangle \cdot 2$ means that the period of coupler population oscillations is multiplied by factor 2, as for the CZ gate two Rabi oscillations of this state are required.}
    \label{fig:rabi_all}
\end{figure}

Going back to the CZ gate realization, first, we should note that our device contains no resonator intended for the readout of the coupling fluxonium, and therefore we cannot measure the coupler directly. To deal with this problem we use the coupler state-dependent frequency shift of the computational qubits. 
Before reading out one of the computational qubits, we apply a low-amplitude, 120~ns-long $\pi$-pulse. The frequency of this pulse corresponds to the qubit frequency when the coupler is unexcited. If the coupler is excited, such an excitation pulse will be out of resonance with the qubit, only weakly affecting the qubit state (see Appendix \ref{appendix:coupler population} for details). 

\begin{figure}[t]
    \center{\includegraphics[width=\linewidth]{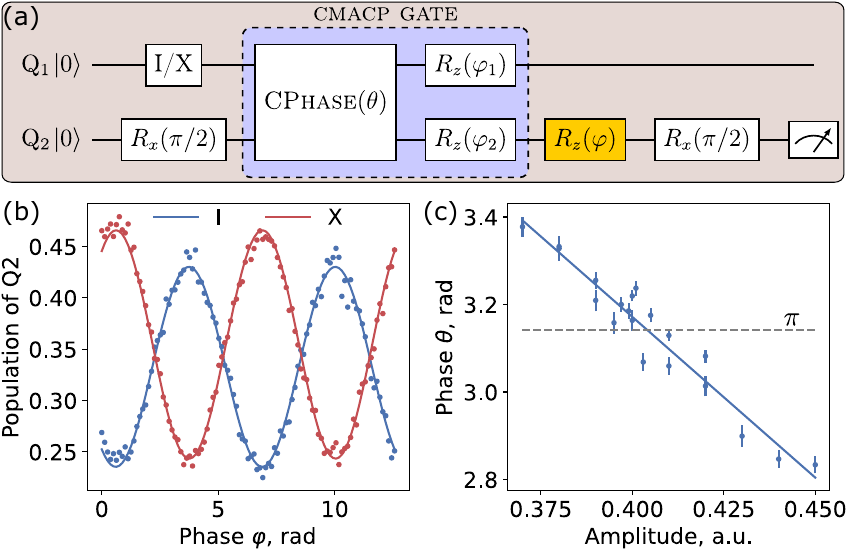}}
    \caption{Calibration of the CZ gate. (a) Excitation pulse sequence. The first qubit is prepared in one of the two states $|0\rangle$ (I gate) or $|1\rangle$ (X gate). Then we apply the coupler microwave-activated controlled-phase gate, make a single-qubit phase rotation on the second qubit and finally measure the second qubit in the X-basis. (b) Populations $p_\mathrm{I}$ and $p_\mathrm{X}$ of the second qubit as a function of the phase $\varphi$ of the highlighted gate. The solid line is the fit of the experimental dots with the function \ref{eq:cz_calibration}. (c)~Conditional phase $\theta$ as a function of the excitation pulse amplitude.}
    \label{fig:cphase_calib}
\end{figure}

\begin{table}[t]
    \centering
    \begin{tabularx}{\columnwidth}{@{}c *4{>{\centering\arraybackslash}X}@{}c}
    \hline
    \hline
         & $\omega_{00}/2\pi$ & $\omega_{10}/2\pi$ & $\omega_{01}/2\pi$ & $\omega_{11}/2\pi$ & \\
         \hline
          value & 3.4098 & 3.44009 & 3.43605 & 3.4660 & GHz\\
          $\sigma$ & 0.1 & 0.07 & 0.08 & 0.2 & MHz\\
    \hline
    \hline
    \end{tabularx}
    \caption{Measured transition frequencies between the ground and excited state of the coupler for the four logical states of the computational qubits $|00\rangle$, $|01\rangle$, $|10\rangle$, and  $|11\rangle$ with corresponding frequencies $\omega_{00}$, $\omega_{01}$, $\omega_{10}$, and $\omega_{11}$. The errors are calculated as a standard deviation of the least squares method.}
    \label{tab:frequencies}
\end{table}

The calibration procedure of the CZ gate starts with determining the transition frequencies of the coupler for the four different states of the computational qubits. For this we perform the pulse sequence shown in~Fig.~\ref{fig:rabi_all}a. We prepare the computational qubits in one of the four initial states $|00\rangle$, $|10\rangle$, $|01\rangle$, $|11\rangle$ and then apply a microwave pulse on the coupler, varying the frequency and duration. 
After that the coupler state measurement protocol consisting of a low-amplitude X-gate applied to one of the qubits followed by the measurement of that qubit is performed.
In this manner, we observe Rabi chevron patterns, from which we extract Rabi frequencies for each detuning of the drive signal. We approximate the data by the formula $\sqrt{\Omega^2+\Delta_{mn}^2}$, where the drive amplitude $\Omega$ and the resonance Rabi frequency $\omega_{mn}$ are the fitting parameters. We collect the results in Table~\ref{tab:frequencies} and plot the drive dependence of the Rabi frequency in Fig~\ref{fig:rabi_all}. 

Using the obtained transition frequencies we calculate the gate duration for an optimal rectangular pulse and experimentally find the corresponding amplitude. Thus we roughly define the frequency, amplitude and duration of the drive pulse. For the precise calibration we execute the gate sequence shown in Fig.~\ref{fig:cphase_calib}, which is commonly used for the phase estimation of a CPhase gate \cite{ganzhorn2020benchmarking}. In this experiment we measure the population of the second qubit as a function of the phase $\varphi$ of the highlighted gate when the first qubit is initially in the $|0\rangle$ and $|1\rangle$ states. These populations are fitted with the functions
\begin{equation}
    \begin{split}
    & p_\mathrm{I}(\varphi) = \frac{1}{2} \left( 1 - \cos (\varphi+\varphi_2) \right) \\
    & p_\mathrm{X}(\varphi) = \frac{1}{2} \left( 1 - \cos (\varphi+\varphi_2+\theta) \right) \\
    \end{split}
    \label{eq:cz_calibration}
\end{equation}
and the phase difference $\theta$ is the parameter of the CPhase gate. We measure this angle for different signal amplitudes, fit the obtained points with a line and find an amplitude corresponding to the $\pi$ phase, see Fig.~\ref{fig:cphase_calib}. Then we measure the period of the Rabi oscillations of the coupler at this amplitude for the computational qubit states $|00\rangle$, $10\rangle$, and $|01\rangle$ (the oscillations associated to the state $|11\rangle$ are indistinct under the current experimental conditions), find the best drive frequency, when these states have close Rabi frequencies and update the gate duration. After that we again refine the amplitude with the previous method. We repeat the procedure in total for three times, until the amplitude converges. 

\begin{figure}[t]
    \center{\includegraphics[width=\linewidth]{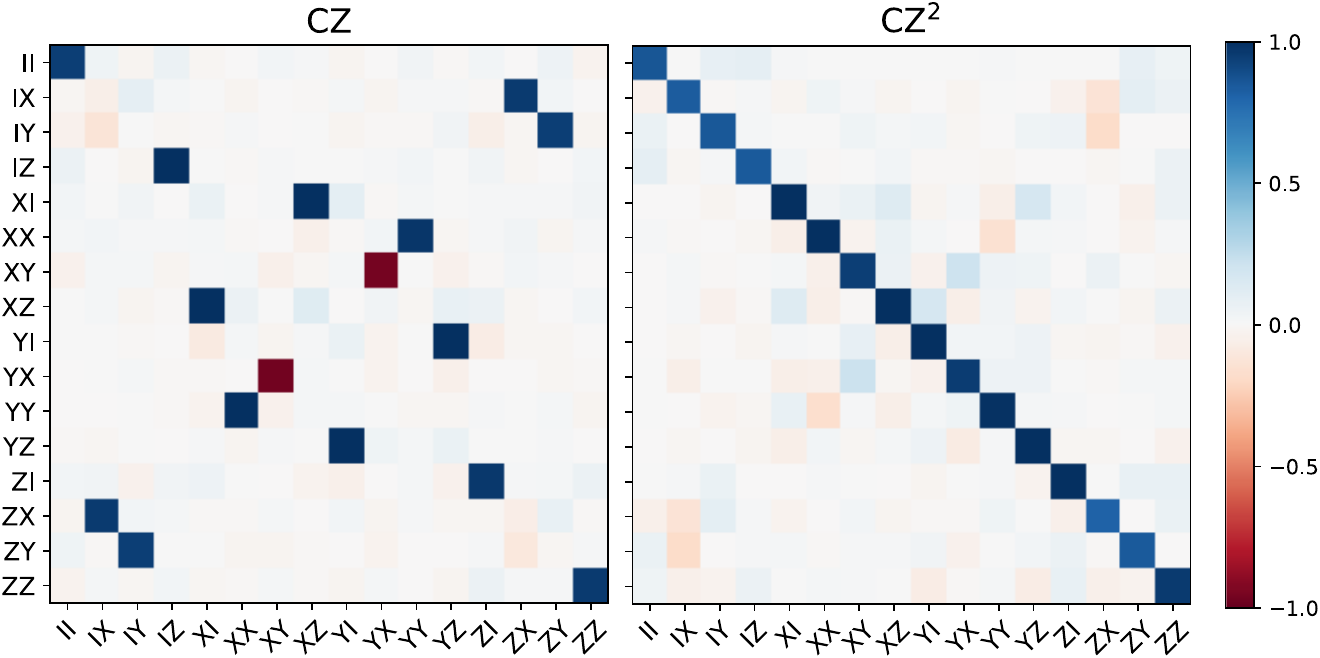}}
    \caption{Experimental reconstruction of the Pauli process matrices for the obtained coupler microwave-activated CZ gate $F=0.9892$ and two CZ gates implemented in a row $F=0.9637$. For comparison the fidelity of the identity matrix is $F=0.9824$.}
    \label{fig:tomography}
\end{figure}

\begin{figure}[t]
    \center{\includegraphics[width=\linewidth]{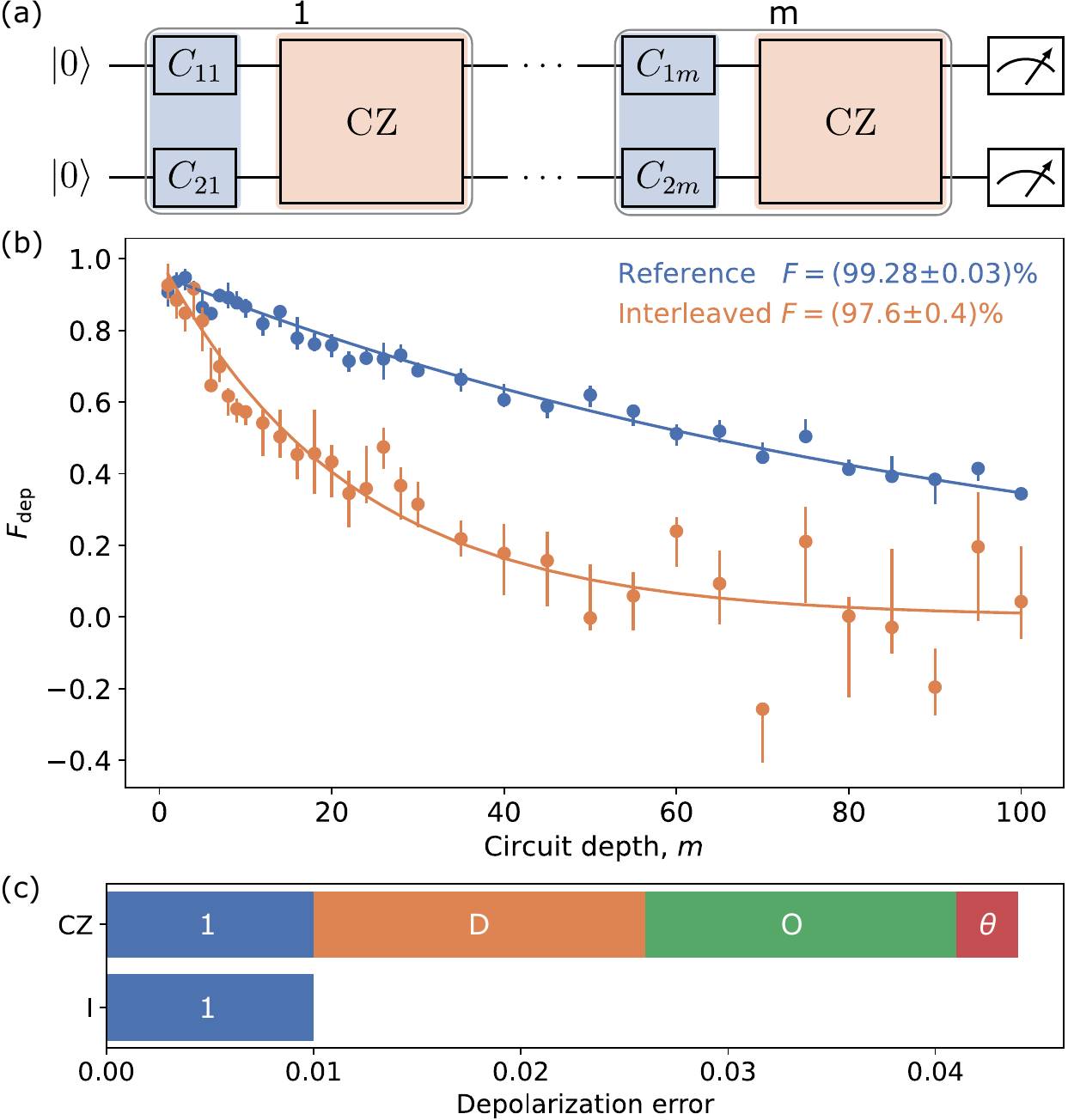}}
    \caption{Cross entropy benchmarking. (a) Quantum circuit for the XEB experiment. (b) Depolarization fidelity for the XEB sequences without (reference) and with interleaved target CZ gate. (c) Depolarization error rate estimates for the implemented gate:  $\varepsilon_\theta$ relates to the non-exact $\pi$ phase of CZ gate; $\varepsilon_1$ and $\varepsilon_\mathrm{D}$ correspond to decoherence on the computational qubits during single-qubit and two-qubit gates, respectively; $\varepsilon_\mathrm{O}$ is the residual error rate which includes effects of residual population and decoherence of the coupling element.}
    \label{fig:xeb}
\end{figure}

Finally, after we determine the parameters of the drive signal, we test the CZ gate. First, we perform a tomography experiment of the two-qubit identity gate I with zero duration, and of the CMACP-generated CZ gate, repeated once and twice. The obtained fidelities are 0.9824, 0.9892 and 0.9637 for the I, CZ and CZ$^2$ gates, correspondingly, and the Pauli process matrices are shown in Fig.~\ref{fig:tomography}. The disadvantage of the tomography method is that the result is strongly affected by state preparation and measurement (SPAM) errors. Thus, in consideration of the poor readout, the fidelity of the CZ gate turned out to be higher than that of the identity gate. The large difference between the fidelities of the CZ and CZ$^2$ gates indicates significant residual population of the coupler after the first CMACP gate, expected for the current system parameters. Indeed, if the coupler does not completely return to the ground state after the gate execution, it does not decrease fidelity of the current gate, but has a strong effect on subsequent gates. 

Second, we test the CZ gate using the cross entropy benchmarking technique (XEB) \cite{arute2019quantum, Boixo2018}. This method accommodates SPAM errors, and is highly sensitive to the residual population of the non-computational states. For the experiment we test quantum circuits with depth $m$ up to 100, for each depth we generate 150 sets of two random sequences $\{C_{11},...,C_{1m}\}$, $\{C_{21},...,C_{2m}\}$ of the single-qubit Clifford gates and average the readout results over 10000 repetitions of each of these sets with and without (reference) the interleaved target CZ gate. We approximate the average depolarization fidelity $F_\mathrm{dep}$ by the function $ap^m$, where $p$ is the depolarization fidelity per cycle and $a$ is a fitting parameter. The conventional gate fidelity can be calculated with the formula
\begin{equation}
    F = p + (1-p)/D,
    \label{eq:fidelity}
\end{equation}
where $D = 2^n$ is the dimension of the Hilbert space ($n=2$). If a target gate is inserted after each single-qubit operation, then the average fidelity of the gate is determined by the eq.~\ref{eq:fidelity} with $p = p_2/p_1$, where $p_2$ and $p_1$ are the depolarization fidelities per cycle corresponding to the interleaved and reference sequences \cite{PhysRevLett123210501}.

In Fig.~\ref{fig:xeb}, blue dots show an exponential decay of the depolarization fidelity of the reference random single-qubit Clifford-gate sequence executed simultaneously on two computational qubits. The orange dots present similar data with the inserted target CZ gate. With a least squares fit we obtain $p_1 = (99.00 \pm 0.05)$\% and $p_2 = (95.6 \pm 0.4)$\%. The average fidelities of the single-qubit Clifford gates and the CMACP-generated CZ gate are  $F = (99.28 \pm 0.03)$\% and $F = (97.6 \pm 0.4)$\%, correspondingly. 

Furthermore, we estimate the contributions of the various error sources to the total error $\varepsilon = 1 - p$ of the CZ gate. The first one is the non-exact conditional phase $\pi$. We find that the fidelity of the XEB data is maximized for the phase of the target gate $\theta = 0.963\pi$ and calculate the depolarization infidelity $\varepsilon_\theta = 0.003$. The next error source is decoherence of the computational qubits. This contribution can be estimated by the extrapolation of the average single-qubit error $\varepsilon_1 = 1-p_1=0.01$ from the duration of a single-qubit operation 26.6~ns to the CZ gate duration 44~ns, yielding $\varepsilon_\mathrm{D} = 0.016$. The rest infidelity $\varepsilon_\mathrm{O} = 0.015$ we attribute to other error sources, that include the decoherence processes and residual population on the coupling fluxonium. 
The estimated error budget is shown with a bar plot in Fig.~\ref{fig:xeb}. The coherence rates and numerical estimations are provided in Appendix~\ref{appendix:coupler population}. Note that the errors for the interleaved sequences are significantly larger than for the reference sequences. This is related to the larger spread of fidelities for different random sequences, which indicates a significant fraction of coherent errors in the implemented two-qubit gates.

\section{Conclusion and outlook}


In this work, we propose and demonstrate a CZ gate realized using a microwave drive applied to a coupler qubit. Due to the strong interaction between the coupling and computational qubits, the main transition frequency of the coupler depends on the state of the computational qubits. This allows to choose such exciting pulse that the speed of phase acquisition on the computational states is different, that in turn results in an effective CPhase operation. We experimentally demonstrate the proposed CZ gate of 0.976 fidelity and 44~ns duration on a device consisting of three capacitively coupled fluxoniums.

The obtained gate fidelity can be improved in further experiments. We identify two significant error sources: decoherence processes and the residual population of the coupler. While decoherence is a significant limiting factor for the performance of many contemporary qubits, the residual population of a degree of freedom mediating the two-qubit interaction is not typical for superconducting qubits. We stress that the latter error has a coherent nature, and can thus be reduced by using advanced control signal shaping \cite{PhysRevApplied.16.024050}.

Both the fluxonium-based topology and the two-qubit gate concept considered in our work demonstrate a promising path forward for scalable and fault-tolerant quantum processors with new qubit types. Tunable couplers in multi-qubit setups help not only to obtain high-fidelity two-qubit operations but to suppress residual XX- and ZZ-coupling rate and avoid frequency collision \cite{Moskalenko_2022}. We also emphasise the low frequency of fluxoniums as a way to simplify individual qubit control via using sub-gigahertz wiring and electronics for gate operations.



\section*{Acknowledgments}

The authors acknowledge Alexey Ustinov for helpful discussions and comments on the manuscript. We acknowledge partial support from the Ministry of Science and Higher Education of the Russian Federation in the framework of the Program of Strategic Academic Leadership “Priority 2030” (MISIS Strategic Project Quantum Internet). Devices were fabricated at the BMSTU Nanofabrication Facility (Functional Micro/Nanosystems, FMNS REC, ID 74300).




\appendix

\section{CZ operation via $\bold{110 - 111}$ transition}
\label{appendix:110-111}

\begin{figure}[t]
    \center{\includegraphics[width=\linewidth]{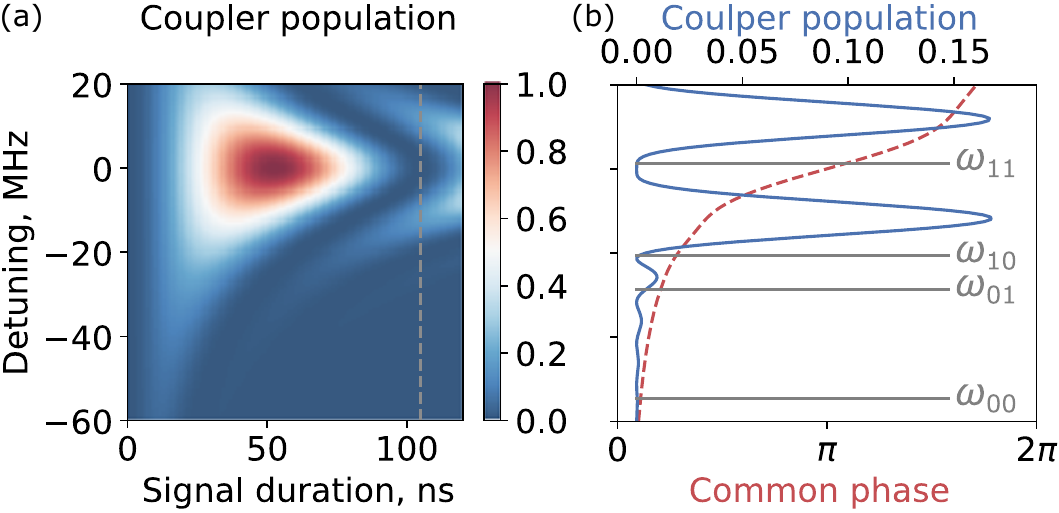}}
    \caption{(a) Rabi oscillations of the coupler population as a function of the exciting pulse frequency and duration. (b) The detuning dependence of the coupler population and the accumulated phase after the first Rabi oscillation, that is shown in the right plot with the grey dashed line. The horizontal lines denote the coupler $0-1$ transition associated with the four computational states. Plot (b) illustrates that the gate requires two conditions: the coupler population should return to the ground state and the accumulated phase of the computational states should satisfy condition~\ref{eq:phase_conditions}.}
    \label{fig:110_111}
\end{figure}

Yet another approach to implement a CZ gate and satisfy condition \ref{eq:phase_conditions} is a direct $2\pi$-pulse drive on the coupling element nearby the resonance $|110\rangle - |111\rangle$ (or $|000\rangle - |001\rangle$).
The idea of the gate is similar to the general parametric CZ operation, implemented via transition between the $|11\rangle - |20\rangle$ states of the computational qubits \cite{ficheux2021fast}. Here, we use the first coupler excited state as an auxiliary energy level out of the computational subspace to acquire conditional phase. 
The major difficulty of the current method is that the signal should not disturb the rest three transitions $|000\rangle - |001\rangle$, $|100\rangle - |101\rangle$ and $|010\rangle - |011\rangle$, thus, we choose the Gaussian envelope for the drive of duration $\tau$:
\begin{equation}
    \Omega(t) = A \left\{ \exp{\left( \cfrac{(t-\tau/2)^2}{2\sigma^2} \right) } -  \exp{\left( \cfrac{(\tau/2)^2}{2\sigma^2} \right) } \right\},
    \label{eq:pulse}
\end{equation}
where $A$ is an amplitude and the pulse length is truncated by $\sigma = 0.4 \tau$. The Rabi pattern caused by this pulse are shown in Fig.~\ref{fig:110_111}a. 

Again, to obtain the CZ operation two conditions have to be satisfied simultaneously. First, at the end of the interaction, the coupler should completely return to the initial state. As it is illustrated in Fig.~\ref{fig:110_111}b, the amplitude of the gate should be picked up in a such way that the coupler population after the gate is low for all possible effective coupler frequencies. 
Here, we search a trade-off between the gate duration and the residual coupler population magnitude.
Thus, the amplitude should be low enough, that the signal do not disturb transition associated with the $|01\rangle$ and $|10\rangle$ computational states, and at the same time high enough to cause a fast Rabi oscillation.
Second, as it can be seen in Fig.~\ref{fig:110_111}b, one cannot neglect the common phase of the computational states $|00\rangle$, $|10\rangle$ and  $|01\rangle$ even if the coupler population associated with them is practically unexcited.
Since the common phase changes more abruptly at the resonance drive than away from it, this condition can be satisfied if one slightly detunes the signal frequency from the $\omega_{11}$.

\begin{figure*}[t]
    \center{\includegraphics[width=\linewidth]{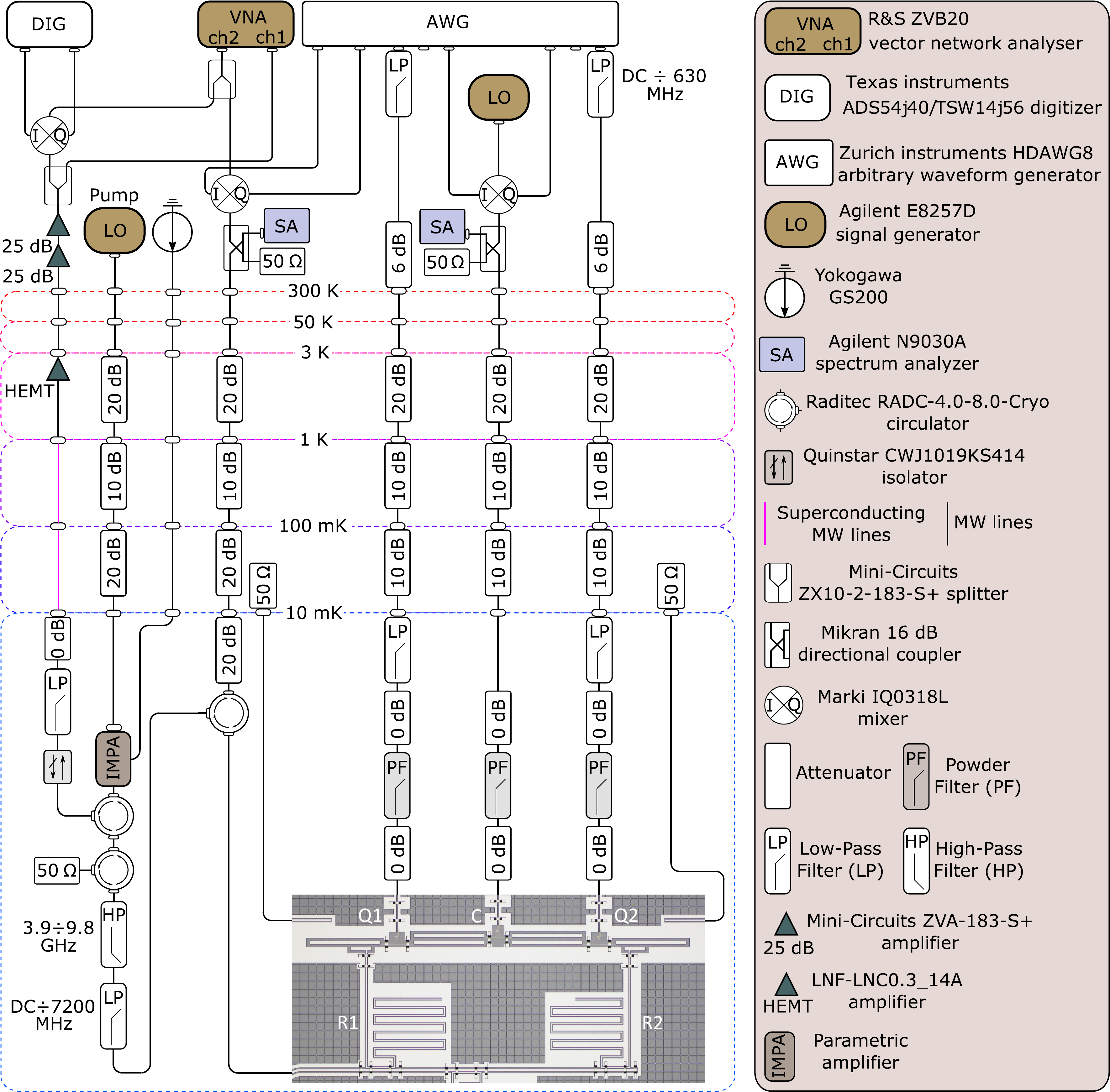}}
    \caption{Experimental setup.}
    \label{fig:scheme_eq}
\end{figure*}

We simulate the time dynamics of the gate by the Schrödinger equation and obtain the gate of 104~ns duration of fidelity 99.9\%. 
The drive frequency is 3.461~GHz and the resonance $|110\rangle - |111\rangle$ frequency is 3.466~GHz. The difference is due to the condition \ref{eq:phase_conditions} and the detuning dependence of the common phase illustrated in Fig.~\ref{fig:110_111}b. This approach has an advantage: it is insensitive to the difference between $\omega_{01}$ and $\omega_{10}$.
However, it is significantly longer than the method used in the main text; hence, it does not give benefits for systems with poor coherence times.

Varying the detuning of the drive signal on can get not only $\pi$, but an arbitrary phase in the expression~\ref{eq:phase_conditions} and consequently obtain a general CPhase gate with arbitrary phase.

\section{Experimental setup}
\label{appendix:exp_setup}

The scheme of the experimental setup is presented in Fig.~\ref{fig:scheme_eq}.
The experiments are performed in a BlueFors LD-250 dilution refrigerator with a base temperature of 10 mK.
The chip is connected to the control setup with six lines: the readout line, two excitation and flux control lines (XYZ controls), two lines coupled with 10 mK stage and ended with 50 Ω terminators for qubit reset, and the coupler’s excitation and flux control line (XYZ controls).

Pulse generation and flux control are fully performed by a Zurich Instruments HDAWG8 arbitrary waveform generator.
One analog output port of the generator is used per fluxonium circuit.
IQ microwave mixers are employed to up- and downconvert the intermediate frequency readout pulses to the resonator frequencies and back.
After getting reflected from the qubit chip, the readout microwave signal is measured by a vector network analyzer (R\&S ZVB20) for spectroscopy and a home-built digitizer setup for single-shot readout.
For mixer calibration we use a spectrum analyzer (Agilent N9030A).

Microwave attenuators are used to isolate the qubit chip from thermal and instrumental noise from the signal sources, which are located at room temperature.
The readout line is equipped with an impedance matched parametric amplifier (IMPA) followed by a Quinstar CWJ1019KS414 isolator to prevent noise from higher temperature stages entering the IMPA and the qubit device.
We pump the IMPA using an Agilent E8257D signal generator.
Three Raditec RADC-4.0-8.0-Cryo circulators and a set of low-pass and high-pass filters placed after the sample preserves it from IMPA pumping and reflected signal. At the PT2 stage (3 K) of the cryostat, a LNF-LNC0.3 14A high electron mobility transistor (HEMT) is installed.
The output line is further amplified outside the cryostat with two Mini-Circuits ZVA-183-S+ amplifiers.
We use a low-pass filter (Mini Circuits VLF-630+) in combination with a powder filter with 15 dB attenuation close to the qubit frequencies in qubit control lines and only powder filter in coupler control line.

Capacitively coupled qubit control lines are connected to 50 Ω terminators at the 10 mK stage of the cryostat.
These lines are used for the qubit initialization and reset.



\section{Device parameters}
\label{appendix:device_parameters}

The current experimental device and a two-qubit processor described earlier in \cite{Moskalenko_2022} are based on the similar design and have been fabricated in the same manufacturing cycles.  
Using the data obtained during the other sample analysis, we get the qubits and coupler $E_C$ and $E_L$ values. With these two parameters fixed, we find the phase slip Josephson junction energy from the frequency at the sweet spot. The device parameters are given in Table \ref{tab:device_parameters}. Also, we provide fundamental transition frequencies $f_{01}$ of each element and coherence times, measured by the standard decay and Ramsey experiments, readout resonator frequencies, and corresponding $\chi_r$ and $\kappa_r$ rates.

\begin{table}[h!]
    \centering
    \begin{tabularx}{\columnwidth}{@{}c *4{>{\centering\arraybackslash}X}@{}c}
    \hline
    \hline
         Parameter & Qubit 1 & Qubit 2 & Coupler \\
         \hline
          $E_C$ (GHz) & 0.55 & 0.55 & 0.584 \\
          $E_L$ (GHz) & 0.7507 & 0.7507 & 0.817 \\
          $E_J$ (GHz) & 1.8 & 1.9 & 2.457 \\
          $f_{01}$ (GHz) & 0.6696 & 0.6944 & 3.4098 \\
          $T_1$ ($\mu$s) & 15.8 & 22.0 & -- \\
          $T_2$ ($\mu$s) & 4.5 & 5.8 & -- \\
          $\omega_{r}/2\pi$ (GHz) & 7.1699 & 7.3813	& --\\
	   $\kappa _{r}/2\pi$ (MHz) & 8.656 & 6.954 & --\\
	   $\chi _{r}/2\pi$ (MHz) & 0.131 & 0.162& --\\
    \hline
    \hline
    \end{tabularx}
    \caption{Device parameters.}
    \label{tab:device_parameters}
\end{table}

\section{Clifford group generation}
\label{appendix:clifford_group}

To use cross-entropy benchmarking technique, one need to generate arbitrary sequences of single-qubit Clifford gates. For an experimenter the task looks like this: you need to perform an arbitrary Clifford gate in the shortest time and make as few preliminary calibration steps as possible. In our previous work~\cite{Moskalenko_2022} we describe in detail the calibration process of single-qubit rotation about X axis at $\pi/2$ angle that we denote as $\sqrt{X}$ gates. Also, we keep in mind that the rotations about Z axis can be virtual and cost no time. Based on this, we propose the following method. We form four sets, each of which contains 24 single-qubit gates:
\begin{equation}
    \begin{split}
        1: \bigl\{ & I, S, S^2, S^3, I, S, S^2, S^3, I, S, S^2, S^3,\\
        & I, S, S^2, S^3, I, S, S^2, S^3, I, S, S^2, S^3 \bigr\}\\
        2: \bigl\{ & \sqrt{X}, \sqrt{X}, \sqrt{X}, \sqrt{X}, \sqrt{X}, \sqrt{X}, \sqrt{X}, \sqrt{X},\\
        & \sqrt{X}, \sqrt{X}, \sqrt{X}, \sqrt{X}, \sqrt{X}, \sqrt{X}, \sqrt{X}, \sqrt{X}\\
        & \sqrt{X}, \sqrt{X}, \sqrt{X}, \sqrt{X}, \sqrt{X}, \sqrt{X}, \sqrt{X}, \sqrt{X} \bigr\}\\
        3: \bigl\{ & I, I, I, I, S, S, S, S, S^2, S^2, S^2, S^2,\\
        & S^3, S^3, S^3, S^3, I, I, I, I, S, S, S, S \bigr\}\\
        4: \bigl\{ & I, I, I, I, I, I, I, I, I, I, I, I, I, I, I, I,\\
        & \sqrt{X}, \sqrt{X}, \sqrt{X}, \sqrt{X}, \sqrt{X}, \sqrt{X}, \sqrt{X}, \sqrt{X} \bigr\}.\\
    \end{split}
\end{equation}
Then we randomly generate an integer number $i$ from 1 to 24 and multiply consecutively the $i$th gate in each set. It is easy to make sure that in this way we get 24 different Clifford operations. Moreover, gates from the first and third sets are virtual, and the gates from the second and fourth sets takes time not exceeding the duration of the $\sqrt{X}$ operation. Thus, implementation of an arbitrary single-qubit Clifford gate reserves a fixed time, that in our case is equal to 26.6 ns.

\section{Coupler state population measurements}
\label{appendix:coupler population}

By the reason of absence of an individual readout resonator for the coupling element, we need a special technique to measure the coupler population. Due to the large interaction between the coupler and the first computational qubit, the data qubit dispersive shift conditioned by the coupler state is significant and equals 26 MHz.
Hence, the drive, optimized for the coupler in the ground state, affects the qubit dynamic differently when the coupler is in the first excited state and the difference is more sufficient for the small amplitude signals.

\begin{figure}[h]
    \center{\includegraphics[width=\linewidth]{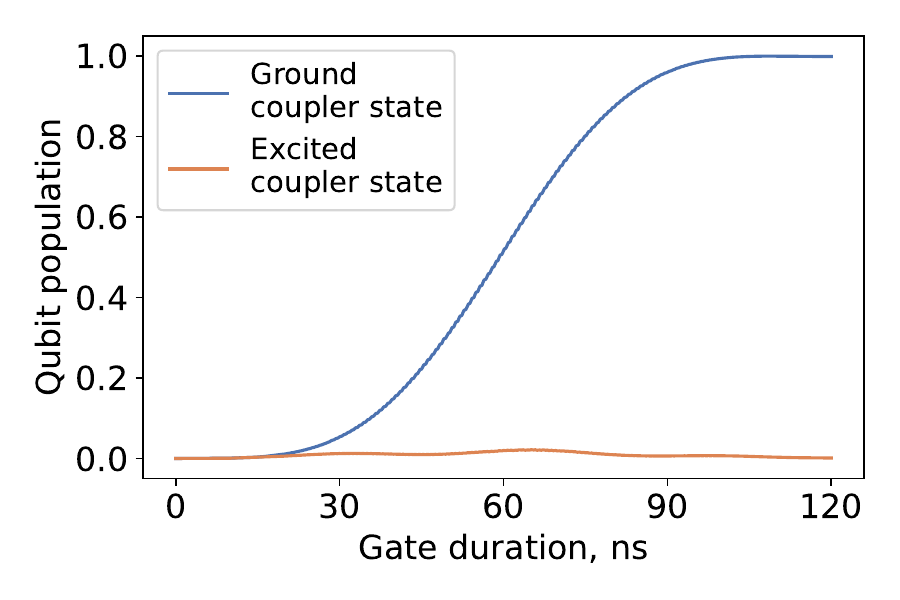}}
    \caption{The first qubit state population dynamic under Gaussian 120-ns $\pi$-pulse for the coupler being in the ground (blue line) and the first excited state (orange line).}
    \label{fig:coupler_meas}
\end{figure}

The idea is as follows, whether the calibrated $\pi$-pulse strongly depends on the coupler state, then we can estimate the coupling element population by its action on the data qubit. In the experiment, we choose the Gaussian $\pi$-pulse with duration 120 ns (the duration of a $\pi/2$-pulse used for Clifford group generation is 13.3 ns). In Fig.~\ref{fig:coupler_meas} we show the numerical simulation of the computational qubit dynamics under this pulse when the coupler is in the ground (blue curve) and excited (orange curve) states.
As one can see, such a long pulse almost preserves the qubit in the ground state for the excited coupler, meanwhile for the unexcited coupler is acts as a high fidelity X gate.


\section{Numerical error estimation}
\label{appendix:residual population}

We consider incoherent and coherent errors. Experimentally obtained coherence times of the computational qubits are $T_1 = 15.8$, $22.0 \; \mu$s and $T_2 = 4.5$, $5.8 \; \mu$s.
To estimate the influence of the decoherence we simulate the evolution of the system with the Lindblad master equation.
The depolarization error of average single-qubit Clifford gates of 26-ns duration is 0.64\%.
The depolarization error of the 44-ns long two-qubit CZ operation simulated under these coherence rates, assuming noiseless coupling qubit, is 1.05\%. 

Due to absence of an individual readout cavity for the coupler the measurements of coupler coherence times were not conducted.
However, as the qubits and the coupler are fluxoniums, we estimate the coupler coherence times at the flux degeneracy point $T_1=18 \; \mu$s and $T_2 = 5 \; \mu$s close to the data qubits rates. The resulted the depolarization error calculated by master equation numerical solution with the noisy coupling element and ideal qubits is 0.28\%.  The discrepancy with experimentally obtained XEB fidelity can be attributed to a shorter coherence time of the coupler, since unlike the computational qubits it is operated at the upper sweet spot.

The residual coupler population affects the fidelity of single-qubit gates implementing after the two-qubit CZ gate.
We compute residual coupler population for each computational state using equation (4). It equals 2.0\% and 1.4\% for $|01\rangle$ and $|10\rangle$ computational states. The residual coupler population of the computational states $|00\rangle$ and $|11\rangle$ is significantly smaller due to the calibration (see right inset plot in Fig 2). 
These populations significantly exceed the equilibrium thermal population of the coupler excited state of 0.07\% for a 3~GHz transition frequency at 20~mK base temperature.
According to the numerical estimation, if the coupler residual population is 2\%, which corresponds to the maximum value after the CZ operation, the single-qubit 26-ns-long X gate fidelity reduces down to 98.7\%.
The XEB technique is sensitive to such error source.
We consider the decrease of the single-qubit gate accuracy as the two-qubit coupler residual population error.

\renewcommand{\bibname}{Reference}
\bibliographystyle{unsrtnat}
\bibliography{main}

\end{document}